\documentclass[sn-mathphys]{sn-jnl}% Math and Physical Sciences Reference Style
\jyear{2022}%

\theoremstyle{thmstyleone}%
\theoremstyle{thmstyletwo}%

\theoremstyle{thmstylethree}%

\begin{document}

\title[Probing the anomalous tqgamma couplings in photon-proton collisions]{Probing the anomalous tqgamma couplings in photon-proton collisions}


\author*[1]{\fnm{E.} \sur{Alici}}\email{edaalici@beun.edu.tr}

\affil*[1]{\orgdiv{Department of Physics}, \orgname{Zonguldak Bulent Ecevit University}, \orgaddress{ \city{Zonguldak}, \postcode{67100}, \country{Turkey}}}


\abstract{The top quark flavor changing neutral current processes are extremely suppressed within the Standard Model. Nevertheless, they could be enhanced in a new physics model beyond the Standard Model. Investigating the top quark's flavor changing neutral current interactions at colliders would be an important test in terms of new physics.
In this work, we examine the potentials of the processes  ${e^{-}} {p} {\rightarrow} {e^{-}} {\gamma} {p}  {\rightarrow}  {e^{-}} W^{+}  b {\gamma}{\rightarrow} {e^{-}} {j} {\bar{j}}  b {\gamma}   $ and ${e^{-}} {p} {\rightarrow} {e^{-}} {\gamma} {p}  {\rightarrow}  {e^{-}} W^{+}  b {\gamma}{\rightarrow} {e^{-}} \textit{l} {\nu}_\textit{l}  b {\gamma}   $    at the FCC-eh with $\sqrt{s}=7.08 $ and $10.0 $ TeV to study anomalous $tq\gamma$ couplings via effective Lagrangians. We obtain $95{\%}$ confidence level sensitivities on the anomalous coupling parameters at the two FCC-eh energies and various integrated luminosities.
We find that our limits are at least two orders of magnitude better than the current LHC experimental results.}

\keywords{FCNC, Top quark physics, photon-proton collisions, FCC}

\maketitle

\section{Introduction}\label{sec1}

The Standard Model (SM) is an excellent theory that defines subatomic particles and their fundamental interactions together. As a result of the discovery of all the foreseen particles by the SM including Higgs boson, it has been proved to be a correct and powerful theory for particle physics \cite{hh,hh1}. However, it can be considered as an effective model of an extended theory owing to some of its serious  deficiencies. In this context, various extended theories which are called beyond the Standard Model (BSM) such as Supersymmetry, Extra Dimensions, Little Higgs and the Standard Model Effective Field Theory (SMEFT) etc. have been proposed. Top quark interactions play a key role and provide very unique exploration opportunities to extended theories due to their huge mass nearly the same amount with the electroweak scale, for the examination of these theories \cite{1}.
A salient feature of the usual top quark sector is that the Flavor Changing Neutral Current (FCNC) transitions are exceedingly weak as a result of the Glashow-Iliopoulos-Maiani (GIM) mechanism. According to the GIM mechanism,  these interactions are prohibited at tree-level and are immensely repressed at the loop-level. The whole of these transitions in the SM are quite beneath the detectable level by experiment, e.g. anticipated branching ratio for decays of $t\rightarrow q\gamma$ $\textit{(q=u,c)}$ are about $10^{-14}$ \cite{2,3}. However, in the BSM theories, the GIM mechanism commonly aren't effective as much as in the SM. Therefore FCNC effects are at a measurable level at the high energy particle colliders. Investigation of these unusual decay channels of the top quark would be an evidence of extended theories \cite{4,5,6,7,8,9,10,11,12}.

In the literature, there are various experimental and phenemological works on FCNC couplings of the top quark to photon and $u$ or $c$ quark via the SMEFT \cite{13,14,15,16,17,18,19,20,21,22,23,24,25,26,27,28,29,30,31,32,33,34,35,36,37,38,39,40,41,42,43,44,45,46,47,48,49,50,51,511}. Up to date, constraints on the branching fractions of top quark FCNC transitions have been achieved by different collaborations \cite{38,39,40,41,42,43,44,45,46,47,48,49,50,51,511}. The most strongest limits related to the anomalous FCNC $tq\gamma$ interactions have been obtained recently by the ATLAS collaborations at $\sqrt{s} = 13$ TeV during Run 2 of the LHC for an integrated luminosity of 139 $fb^{-1}$ \cite{511}. These experimental constraints on the branching fraction, via left-handed (right handed) tq$\gamma$ couplings have been achieved as $BR(t\rightarrow u\gamma)<0.85\times10^{-5}(1.2\times10^{-5})$ and $BR(t\rightarrow c\gamma)<4.2\times10^{-5}(4.5\times10^{-5})$ \cite{511}. Despite many experimental researches, there haven't been found any hint for anomalous FCNC transitions of the top quark, at the present colliders. Therefore, future particle colliders, which are planned to include new technologies and different type of collisions from present accelerators, are gaining importance in the FCNC researches.

The LHC, the available most powerful and highest resolution collider, will complete its mission in the 2030s. According to the conceptual design report (CDR) revealed by CERN in 2012 and revised in 2017, a new collider called Future Circular Collider (FCC) will be built in the post-LHC area. FCC is designed as a proton-proton collider with a 100 TeV centre-of mass energy \cite{52,53,saleh,54}. Furthermore, the FCC will have different interaction possibilities. Electron-proton and photon-proton interactions will be possible thanks to a special electron linear collider (ERL) that will be built tangentially to the ring-shaped main tunnel of the FCC. That collider is called FCC-eh has three different centre-of mass energy options. In this investigation, two of them have been taken into account as 7.08 TeV and 10.0 TeV. Here, the energies of their electron beam are 250 GeV and 500 GeV, respectively, while the energies of the proton beam are 50 TeV for both of them. Although electron-proton and photon-proton collider type have lower centre-of mass energy than proton-proton collider,  they allow a clean background free from the pile-up and multiple interactions that are formed as a result of strong interaction in proton-proton processes. Hence, they are of great importance in order to examine the main structure of matter.

In this paper, the photon-proton processes, in which the electron formed a photon, are taken into consideration. The photons, which are named quasi-real photons, arise from spontaneous radiation of the ERL electron beam. The electron beam moving in the special linac loses some of its transverse momenta and emits  photons which scatter from the direction of movement at a very small angle. These photons are consistent with Weizsacker-Williams approximation (WWA) \cite{55,56,57}.

\section{Theoretical Formalism}\label{sec2}
FCNC transitions can be possible via the Standard Model Effective Field Theory (SMEFT) depend on effective low energy lagrangian which has the same symmetry group with the SM. With the new terms added to the SM lagrangian, effective lagrangian is defined as the following form \cite{59,60},
\begin{eqnarray}
{\mathcal{L}}={\mathcal{L}}_{\textit{SM}}+\sum\frac{C_{i}O_{i}}{{\Lambda}^2}.
\end{eqnarray}
Here, ${\mathcal{L}}_{\textit{SM}}$ defines the SM Lagrangian that contains four dimensional operators while $O_{i}$ describe six dimensional operators related to extended theories. In addition, ${\Lambda}$ indicates the energy scale of extended theories and $C_{i}$ identify the Wilson couplings. In this paper, we have decided to use the terms of the new physics which inculude the dimension six operators describing FCNC couplings among the top quark, q(u quark and c quark) and photon \cite{59},
\begin{eqnarray}
\begin{split}
O^{ij}_{uW}=(\bar{q}_{Li}{\sigma^{\mu\nu} }{\tau}^{I}{u}_{Rj})\tilde{\phi}{}W^{I}_{{\mu}{\nu}} ,\\
O^{ij}_{uB\phi}=(\bar{q}_{Li}{\sigma^{\mu\nu} }{u}_{Rj})\tilde{\phi}{}B_{{\mu}{\nu}},
\end{split}
\end{eqnarray}
where i, j are flavour indices.  ${q}_{Li}$ is left-handed quark doublet while ${u}_{RJ}$ is right-handed quark singlet. ${\tau}^{I}$ defines the Pauli matrices. Furthermore, $\tilde{\phi}=i{\tau}^{2}\phi^{*}$, where  the $\phi$ corresponds SM Higgs doublet.
The SM Lagrangian can be extended easily with the FCNC couplings in the vertex of tq${\gamma}$ as follows \cite{581,59},	
\begin{eqnarray}
{\mathcal{L}_{FCNC}}=\frac{g_{e}}{2m_{t}}\sum_{q=u,c}\bar{q}{\sigma_{\mu\nu} }({\lambda}_{qt}^{R}P_{R}+{\lambda}_{qt}^{L}P_{L})t A^{{\mu}{\nu}}+h.c .
\end{eqnarray}
In the above equation, $\sigma_{\mu\nu}=[\gamma_{\mu},\gamma_{\nu}]/2  $ , $g_{e}$ symbolizes the electromagnetic coupling constant. Furthermore, $P_{L}$ and $P_{R}$ determine the left and right-handed projection operators, respectively. $ {\lambda}_{qt}^{R(L)}$ are dimensionless real parameters, which include the Wilson coefficients of effective operators and stand for the anomalous couplings constant between a top quark, u quark or c quark and a photon. The anomalous couplings are related to the dimension six operators as,
\begin{eqnarray}
\begin{split}
\lambda^{L}_{qt}=\frac{\sqrt{2}}{e}[s_{w}C^{3j*}_{uW}+c_{w} C^{3j*}_{uB{\phi}}]\frac{{\nu} m_{t}}{{\Lambda}^2},\\
\lambda^{R}_{qt}=\frac{\sqrt{2}}{e}[s_{w}C^{j3}_{uW}+c_{w} C^{j3}_{uB{\phi}}]\frac{{\nu} m_{t}}{{\Lambda}^2} .
\end{split}
\end{eqnarray}
In the presented study, it was assumed there is no specific chirality for the FCNC interaction vertices to simplify scenario, i.e. $ {\lambda}_{qt}^{R}={\lambda}_{qt}^{L}={\lambda}_{q}$. Additionally, $ \lambda_{ut}$ and  $ \lambda_{uc}$ have been considered as equal \cite{28,31,582}. Decay width of top decaying to $q\gamma$ through  tq${\gamma}$ coupling  can easily obtained as ${\Gamma(t{\rightarrow}q\gamma)}=\frac{\alpha}{2}{\lambda_{q}^2 } m_t {\backsimeq} {0.6528{\lambda_{q}^2 }}$ by utilizing Eq(2). Here, $\alpha$ is the fine structure constant related to ge as $\alpha=(g_{e}e)^2/(4\pi)=\frac{1}{132.1}$ and $m_t=173.0$ GeV. As known, the branching ratio for the anomalous coupling interaction is given as follows,
  \begin{eqnarray}
  BR(t\rightarrow q\gamma)=\frac{\Gamma(t\rightarrow q \gamma)}{\Gamma(t\rightarrow Total)} .
  \end{eqnarray}
However, in the SM, the primary decay channel of the top quark sector is the mode which a W boson and a bottom quark are produced. Accordingly, ${\Gamma(t\rightarrow W b)}$ decay width can be substituted instead of the top quark total decay width in the Equation (5). The decay width of this channel is approximately 1.49 GeV. Both FCNC decay widths ($\Gamma(t\rightarrow q \gamma)$) and total decay width ($\Gamma(t\rightarrow W b)$) of the top quark are computed by MadGraph5$\_$aMC@NLO \cite{61}. Based on the calculated data, the branching ratio of anomalous couplings can be  written as fraction of $t\rightarrow q\gamma$ $\textit{(q=u,c)}$ decay width to $t\rightarrow W b$ decay width ($BR(t\rightarrow q\gamma)=\frac{\Gamma(t\rightarrow q \gamma)}{\Gamma(t\rightarrow W b)}$). Consequently, it can be easily formulized as $BR(t\rightarrow q\gamma)={0.4573{\lambda_{q}^2}}$.
\subsection{Cross Sections}\label{sec3}
In this paper, we have examined ${e^{-}} {p} {\rightarrow} {e^{-}} {\gamma}{p}  {\rightarrow}  {e^{-}} W^{+}  b {\gamma}{\rightarrow} {e^{-}} {j} {\bar{j}}  b {\gamma}$ (j= u,d,b,c,s and $\bar{j}=\bar{u},\bar{d},\bar{b},\bar{c},\bar{s}$) and ${e^{-}} {p} {\rightarrow} {e^{-}} {\gamma} {p}  {\rightarrow}  {e^{-}} W^{+}  b {\gamma}{\rightarrow} {e^{-}} {\textit{l}} {{\nu}_\textit{l}}  b {\gamma} ,(\textit{l}=e, {\mu}$ and ${\nu}_\textit{l}={\nu}_{\textit{e},{\mu}} ) $,  processes via the SMEFT. The tree level Feynmann diagrams for the subprocesses $ \gamma q (q=u,c) {\rightarrow}W^{+}  b {\gamma}(W^{+}{\rightarrow}{j} {\bar{j}},{\textit{l}} {{\nu}_\textit{l}})$, that include new physics contributions are shown in Fig. 1. In the figures and tables,
${e^{-}} {p} {\rightarrow} {e^{-}} {\gamma}{p}$ $ {\rightarrow}{e^{-}} W^{+}  b {\gamma}{\rightarrow} {e^{-}} \textit{l} {\nu}_\textit{l}  b {\gamma}   $ processes are mentioned as leptonic channel and ${e^{-}} {p} {\rightarrow} {e^{-}} {\gamma} {p}  {\rightarrow}  {e^{-}} W^{+}  b {\gamma}$ $  {\rightarrow} {e^{-}} {j} {\bar{j}}  b {\gamma}  $ as hadronic channel. To generate these processes, we utilized   the Monte Carlo simulation program MadGraph5$\_$aMC@NLO \cite{61}.  We expanded The FeynRules package \cite{62} , in which the Standard Model lagrangian is defined, by adding Universal FeynRules Output (UFO) modul \cite{612} including SMEFT lagrangian terms \cite{581}.  In this UFO modul, SM-ckm matrix is diagonal. In order to obtain the SM background of the process $ \gamma q {\rightarrow}W^{+}  b {\gamma},(W^{+}{\rightarrow}{j} {\bar{j}},{\textit{l}} {{\nu}_\textit{l}})$, the transitions from top and charm quark to bottom quarks must be allowed. Therefore, non-diagonal ckm matrix elements were embedded into the program.

\begin{figure}[h]
\centering
\includegraphics[width=0.9\textwidth]{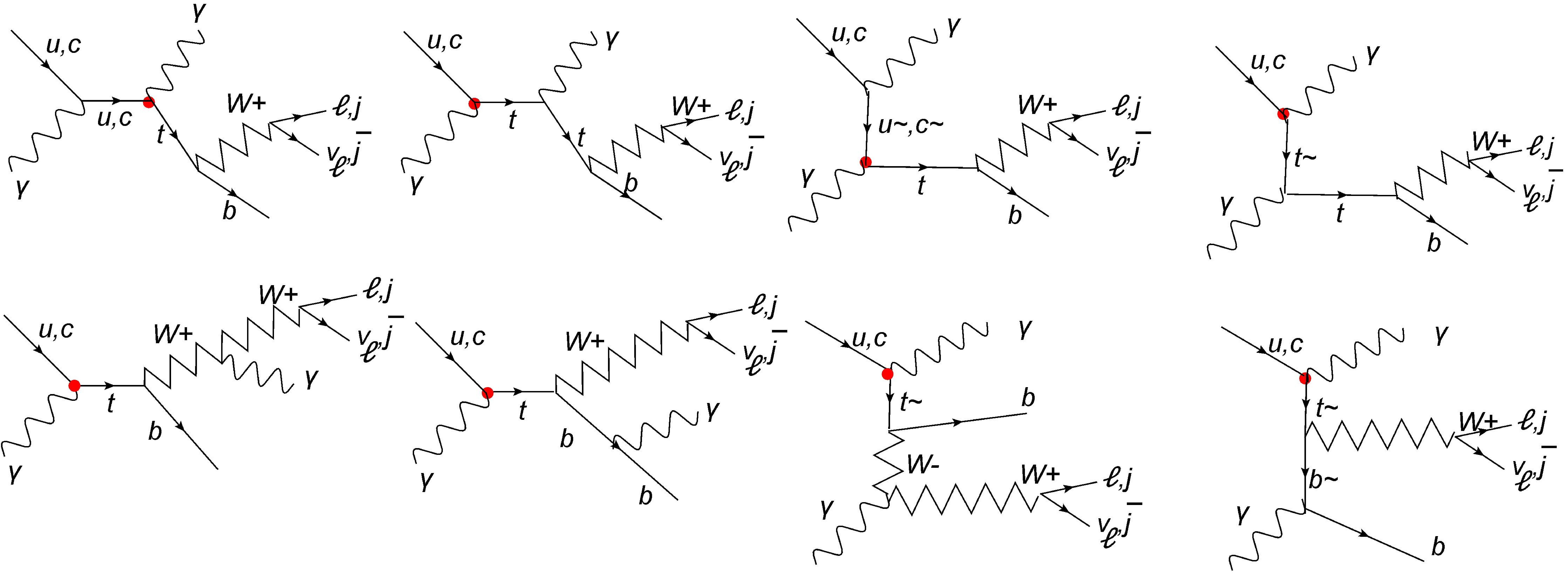}
\caption{ Tree level Feynman diagrams for the subprocesses $ \gamma q {\rightarrow}W^{+}  b {\gamma},(W^{+}{\rightarrow}{j} {\bar{j}},{\textit{l}}{} {{\nu}_\textit{l}}) (q=u,c) $ that includes new physics contributions}
\end{figure}
For analysis, we have chosen the CTEQ6L1 parton distribution function because it also has WW photon distribution function \cite{623}. For the $ {\gamma} q $ processes, the distribution function of WW photon is given by following equation \cite{55,56,57,58},
\begin{eqnarray}
{f_{{\gamma}}(x)=\frac{\alpha}{\pi E_{e}}\{[\frac{1-x+x^{2}}{x}]log(\frac{Q_{max} ^ {2}}{Q_{min} ^ {2}})-\frac{m_{e}^{2}x}{Q_{min} ^ {2}}(1-\frac{Q_{max} ^ {2}}{Q_{min} ^ {2}})-\frac{1}{x}[1-\frac{x}{2}]^{2}log(\frac{x^{2}E_{e}^{2}+Q_{max} ^ {2}}{x^{2}E_{e}^{2}+Q_{min} ^ {2}})\}} ,
\end{eqnarray}
where $x=\frac{E_{{\gamma}}}{E_{e}}$ and $Q_{max}^{2}$ is the maximum virtuality of the photon. During the calculation, the photon virtuality is taken as $Q_{max} ^ {2} = 2$ GeV$^{2}$. The minimum value of $Q_{min}^{2}$ is given as follows,
\begin{eqnarray}
{Q_{min} ^ {2}=\frac{m_{e}^{2}x}{1-x}}.
\end{eqnarray}

For the processes in this study, we applied different cuts to distinguish the SM cross section from the new physics. For this purpose, we first applied   for $p_{t}^{\textit{l},b,j ,\bar{j}}> 20$ GeV, $ MET> 20$ GeV and $ -2.5<{\eta}^{\textit{l},b,j ,\bar{j},{\gamma}}<2.5 $ the cuts named cut1. Here, pt is the transverse momenta of the relevant particles while $\eta$ is the pseudorapidty. As well as, different cuts can be chosen in addition to cut1 to separate the new physics signal from the SM background signal. In this regard, the effect of the $p_{t}^{\gamma}$ variable distribution of the photon which is the final state particle on the number of events can be examined. In this paper,  we achieved the relevant investigations using Madanalysis5 \cite{624}. Accordingly, Figure 2 and 3 shows the variation of the number of events in the $0-500$ GeV value range of $p_{t}^{\gamma}$. Here, it has been accepted as $\mathcal{L}= 10 fb^{-1}$ while examining the number of events. Here, cut1 and ${\lambda}_q=0.01$ selections are applied in the calculations. Figure 2 and 3 depicted for center-of mass energy of $\sqrt{s}=7.08$ TeV and $\sqrt{s}=10.0$ TeV, respectively and  the top panels in the figures represent the leptonic channel while the bottom ones represent the hadronic channel. When the Fig2 and 3 were examined, it has been realized that the $p_{t}^{\gamma}$ distribution of the total number of events containing new physics contributions is remarkably different from the $p_{t}^{\gamma}$ distribution of the number of SM events. Moreover, according the figures, we can say that the $p_{t}^{\gamma}$ -dependent behavior of the SM events and the total events begin to differ significantly from each other after $p_{t}^{\gamma}=50$ GeV .
\begin{figure*}[h]
    \centering
    \includegraphics[width=10cm]{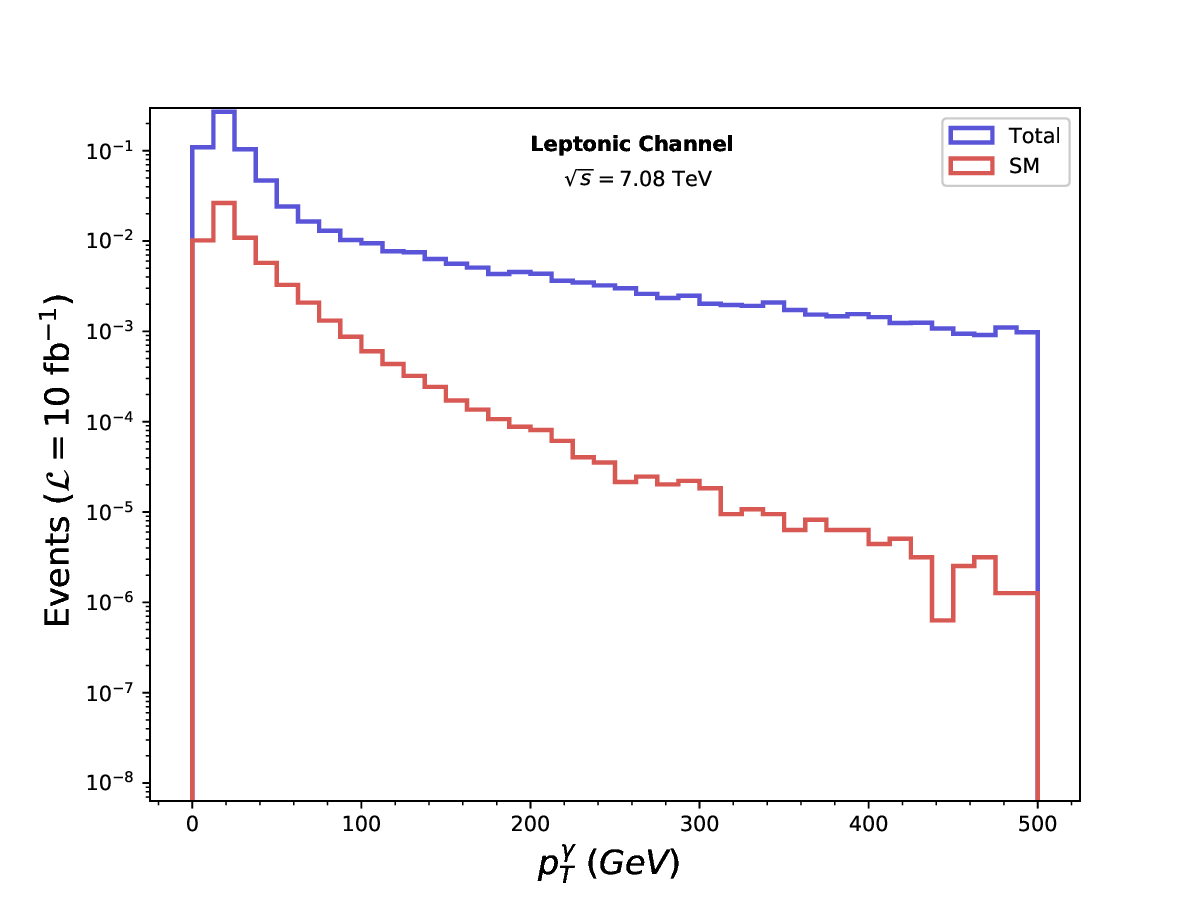}
    \includegraphics[width=10cm]{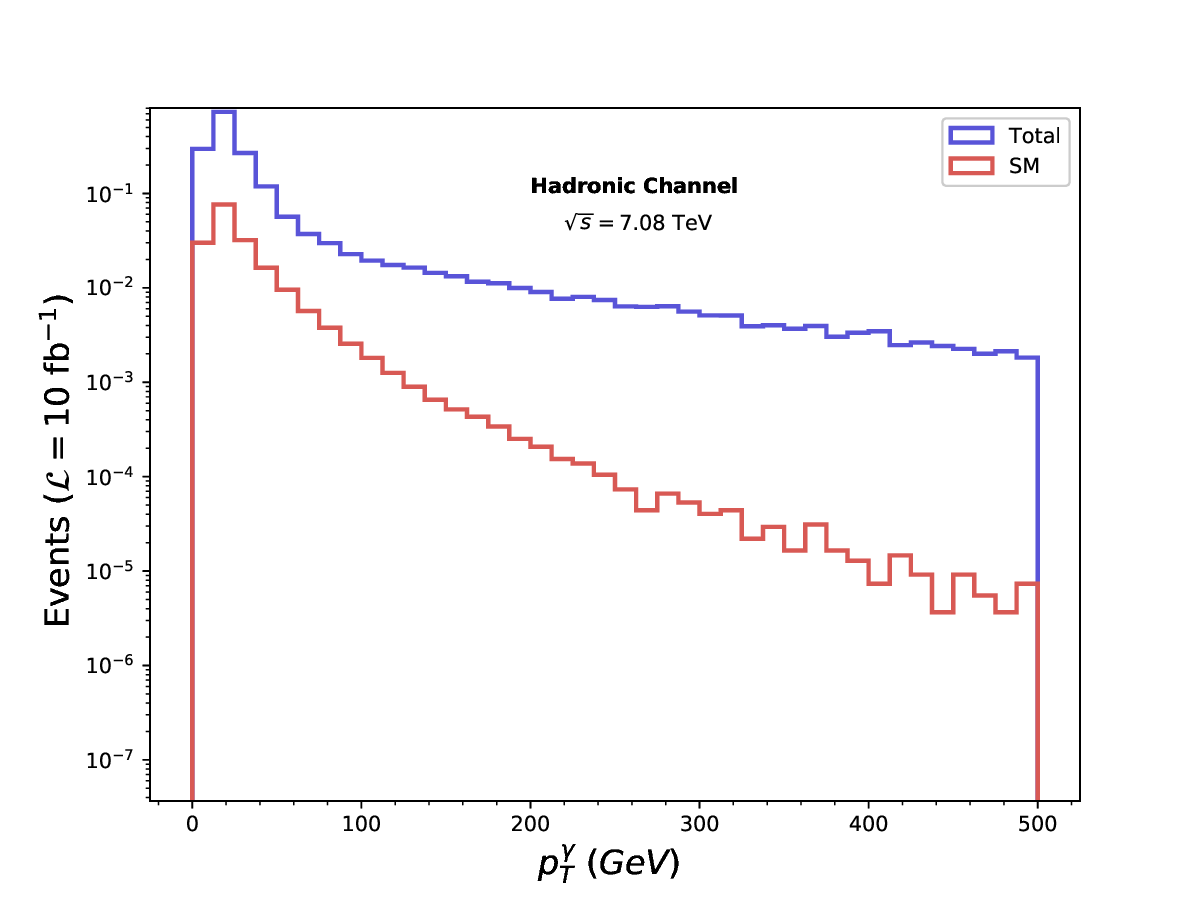}
    \caption{The $p_{t}^{\gamma}$ distributions of the total events and SM events at the center of mass energy of $\sqrt{s}=7.08 $ TeV }%
    \label{fig:example}%
\end{figure*}

\begin{figure*}[h]
    \centering
    \includegraphics[width=10cm]{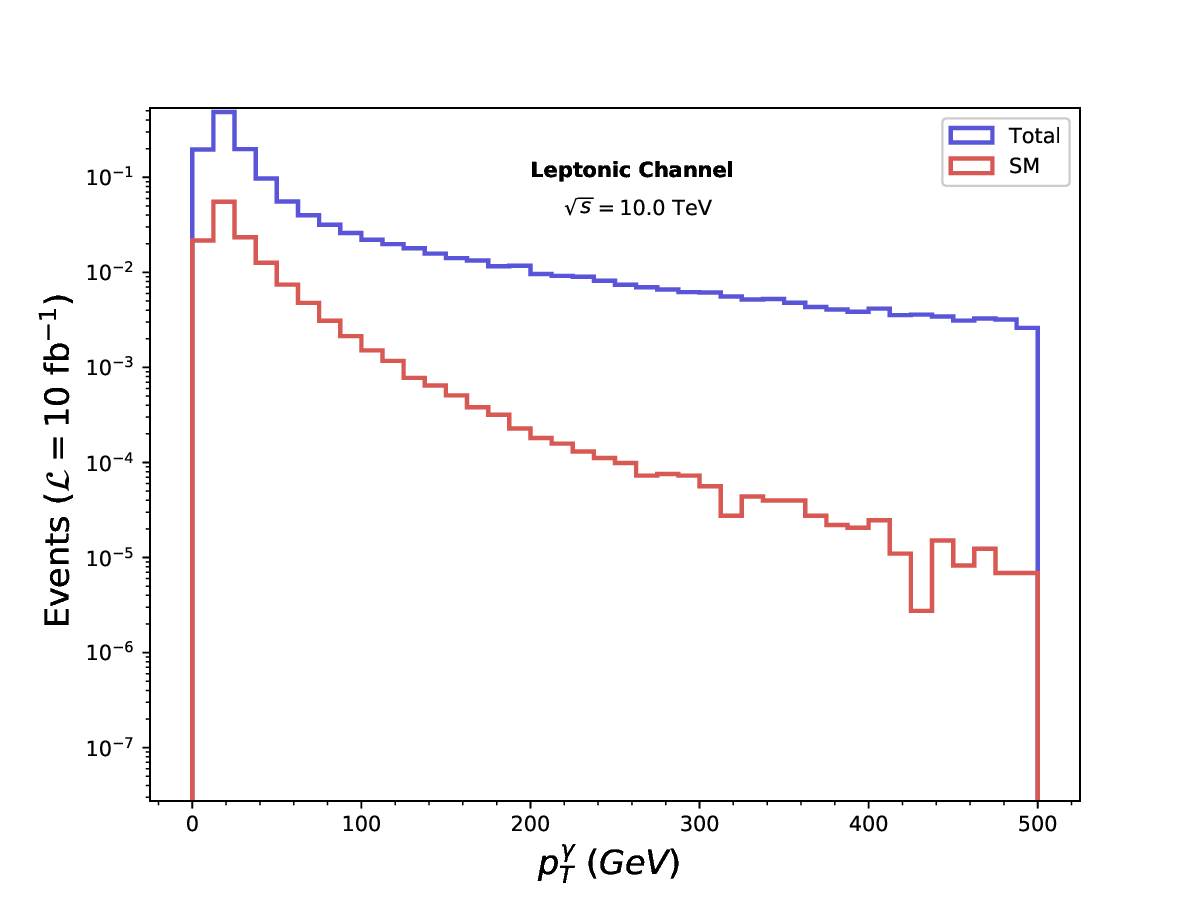}
    \includegraphics[width=10cm]{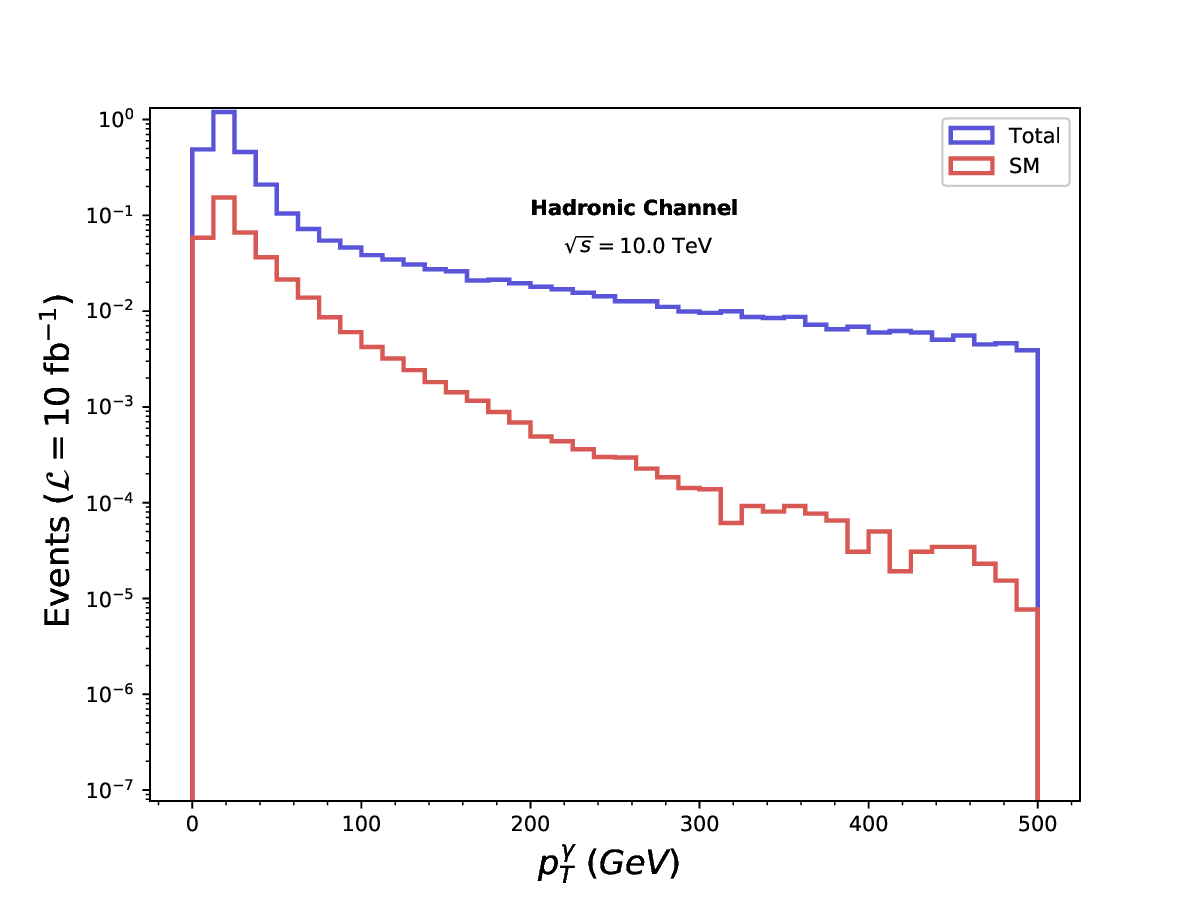}
    \caption{The $p_{t}^{\gamma}$ distributions of the total events and SM events at the center of mass energy of $\sqrt{s}=7.08 $ TeV }%
    \label{fig:example}%
\end{figure*}
Furthermore, in order to examine the $p_{t}^{\gamma}$ effect between 0-50 GeV in more detail, the four different transverse momenta values of the photon $p_{t}^{\gamma}>20 $ GeV, $p_{t}^{\gamma}>30$ GeV, $p_{t}^{\gamma}>40$ GeV and $p_{t}^{\gamma}>50$ GeV have been selected together with cut1, and these cuts are named as cut2, cut3, cut4 and cut5, respectively. By applying these different cuts, we investigated the suppressing effect of the photon's transverse momenta on the SM cross section. The relevant cuts are given in Table 1.
\begin{table}[h]
\begin{center}
\caption{List of applied kinematic cuts for the analysis}
\begin{tabular}{cccc|cccc}
\toprule
Cut1& \hspace{0.5cm} $p_{t}^{\textit{l},{{\nu}_\textit{l}},b,j ,\bar{j}}> 20$ GeV, MET $> 20$ GeV ,$ -2.5<{\eta}^{\textit{l},b,j ,\bar{j},{\gamma}}<2.5 $ &\\
Cut2&   Cut1+$p_{t}^{\gamma}>20$ GeV & \\
Cut3& Cut1+$p_{t}^{\gamma}>30 $  GeV    & \\
Cut4&  Cut1+$p_{t}^{\gamma}>40$ GeV &  \\
Cut5&  Cut1+$p_{t}^{\gamma}>50$ GeV & \\
\botrule
\end{tabular}
\end{center}
\end{table}

In this regard, SM cross section values were calculated and the ratios of the obtained these values to the total cross section values calculated for the new physics parameter ($\lambda_q=0.01$) value were found. Thus, we determined which of the used cuts would include (make visible) new physics effects. The obtained ratio and cross section values for both leptonic and hadronic channels were given for two different centre of mass energies 7.08 TeV and 10.0 TeV in Table II and Table III, respectively. As it can be seen from the calculations at the $\sqrt{s}=7.08 $ TeV in Table II, the SM cross sections for the $ \gamma q {\rightarrow}W^{+}  b {\gamma}{\rightarrow}{\textit{l}} {{\nu}_\textit{l}}b {\gamma}$ processes (leptonic channel) is greatly suppressed by the cut1 setup and reduced the order of $10^{-6}$ pb. Here, it is, also, observed that the SM cross section decreases the order of $10^{-7}$ pb with the cut5 setup. On the other hand, when the SM cross section values obtained for the $ \gamma q {\rightarrow}W^{+}  b {\gamma}{\rightarrow}{j} \bar{j}b {\gamma}$ subprocesses (hadronic channel) in Table II are examined, a very small SM cross section in the order of $10^{-6}$ pb is obtained in the cut2 setup. For the calculations at the $\sqrt{s}=10.0 $ TeV in Table III, it can be said that the SM cross-section value in the order of $10^{-6}$ pb is obtained for the cut2 setup in the leptonic channel and for the cut4 setup in the hadronic channel. Moreover, the high ratio values in the tables correspond to the high suppression effect on the SM background. Evaluating Table II and Table III together, it was seen that there are almost no difference in the obtained ratio values for cut1 and cut2. This can be explained by the fact that a minimum of $p_{t}^{\gamma}>20$ GeV cuts put into the transverse momenta of the photon between the cuts do not suppress the SM contributions. On the other hand, examining the ratio values from cut2 to cut5, it was seen that the most striking increase was in cut5, which was put a $p_{t}^{\gamma}>50$ GeV in addition to cut1.

\begin{table}[h]
\caption{For five different cuts, total and SM cross section values and the ratios of total cross section to SM cross
section on the anomalous $tq\gamma$ couplings at the center of mass energy of $\sqrt{s}=7.08 $ TeV}
\begin{center}
\begin{tabular}{cc|cc|cc|cc|cc}
\toprule
\multicolumn{10}{c}{$\sqrt{s}=7.08 $ TeV} \\
\midrule
\multicolumn{2}{c|}{}& \multicolumn{2}{c|}{Cuts} & \multicolumn{2}{c|}{$\sigma_{SM}$(pb)} & \multicolumn{2}{c|}{$\sigma_{Tot}$(pb)}& \multicolumn{2}{c}{Ratio} \\
\midrule
&& cut1&& $6.257\times10^{-6}$   &&$6.55\times10^{-5}$&& 10.4  & \\
&&cut2&& $3.389\times10^{-6}$   &&$3.90\times10^{-5}$&& 11.5  &   \\
Leptonic Channel&&cut3&&$2.125\times10^{-6}$   &&$2.748\times10^{-5}$&& 12.9  &  \\
&&cut4&&$1.396\times10^{-6}$   &&$2.083\times10^{-5}$&& 14.9 &  \\
&&cut5&&$9.839\times10^{-7}$  &&$1.79\times10^{-5}$&& 18.2  &  \\
\midrule
&&cut1&&$1.826\times10^{-5}$   &&$1.758\times10^{-4}$&& 9.6  & \\
&&cut2&&$6.105\times10^{-6}$  &&$6.262\times10^{-5}$&& 9.7  &   \\
Hadronic Channel  &&cut3&&$1.37\times10^{-6}$   &&$5.33\times10^{-5}$&& 10.2  &  \\
&&cut4&&$4.113\times10^{-6}$   &&$4.745\times10^{-5}$&& 11.5  &  \\
&&cut5&&$2.826\times10^{-6}$   &&$3.952\times10^{-5}$&& 14.0  &  \\
\botrule
\end{tabular}
\end{center}
\end{table}

\begin{table}[h]
\caption{For five different cuts, total and SM cross section values and the ratios of total cross section to SM cross
section on the anomalous $tq\gamma$ couplings at the center of mass energy of $\sqrt{s}=10.0 $ TeV}
\begin{center}
\begin{tabular}{cc|cc|cc|cc|cc}
\toprule
\multicolumn{10}{c}{$\sqrt{s}=10.0 $ TeV} \\
\midrule
\multicolumn{2}{c|}{}& \multicolumn{2}{c|}{Cuts} & \multicolumn{2}{c|}{$\sigma_{SM}$(pb)} & \multicolumn{2}{c|}{$\sigma_{Tot}$(pb)}& \multicolumn{2}{c}{Ratio} \\
\midrule
&& cut1&& $1.354\times10^{-5}$   &&$1.406\times10^{-4}$&& 10.4  & \\
&&cut2&& $7.418\times10^{-6}$   &&$8.748\times10^{-5}$&& 11.8  &   \\
Leptonic Channel&&cut3&&$4.854\times10^{-6}$   &&$6.573\times10^{-5}$&& 13.5  &  \\
&&cut4&&$3.252\times10^{-6}$   &&$5.287\times10^{-5}$&& 16.3 &  \\
&&cut5&&$2.382\times10^{-6}$  &&$4.752\times10^{-5}$&& 19.9  &  \\
\midrule
&&cut1&&$3.803\times10^{-5}$   &&$3.07\times10^{-4}$&& 8.0  & \\
&&cut2&&$2.134\times10^{-5}$  &&$1.704\times10^{-4}$&& 8.0  &   \\
Hadronic Channel  &&cut3&&$1.373\times10^{-5}$   &&$1.168\times10^{-4}$&& 8.5  &  \\
&&cut4&&$9.615\times10^{-6}$   &&$9.187\times10^{-5}$&& 9.5  &  \\
&&cut5&&$6.825\times10^{-6}$   &&$7.841\times10^{-5}$&& 11.4  &  \\
\botrule
\end{tabular}
\end{center}
\end{table}

Considering all these findings, cut5 are used for the rest of our study. The dependence of the cross section of the
${e^{-}} {p} {\rightarrow} {e^{-}} {\gamma} {p}  {\rightarrow}  {e^{-}} W^{+}  b {\gamma}{\rightarrow} {e^{-}} {j} {\bar{j}}  b {\gamma}   $ and ${e^{-}} {p} {\rightarrow} {e^{-}} {\gamma} {p}  {\rightarrow}  {e^{-}} W^{+}  b {\gamma}{\rightarrow} {e^{-}} \textit{l} {\nu}_\textit{l}  b {\gamma}   $   subprocesses on the new physics parameter ${\lambda}_q$ for $\sqrt{s}=7.08 $ TeV and $\sqrt{s}=10.0 $ TeV is depicted in Figure 4 and Figure 5, respectively. As it is realized from the figures, the cross-section values of hadronic processes are higher than ones of leptonic processes. Comparing the figures for the same anomalous coupling constant, it was observed that the cross section is also high when the centre-of-mass energy is high. In fact, this is an expected result, since the new physics terms contain high energy dependence. In addition, new physics contributions along with the increased anomalous coupling constant have an effect far above the SM cross section.
\begin{figure}[h]
  \centering
  \includegraphics[width=10.cm]{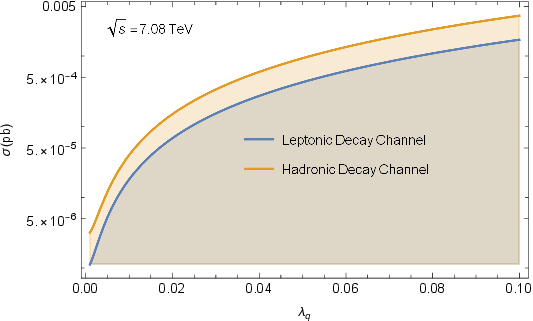}
   \caption{The total cross sections of the processes ${e^{-}} {p} {\rightarrow} {e^{-}} {\gamma} {p}  {\rightarrow}  {e^{-}} W^{+}  b {\gamma}{\rightarrow} {e^{-}} {j} {\bar{j}}  b {\gamma}   $ and ${e^{-}} {p} {\rightarrow} {e^{-}} {\gamma} {p}  {\rightarrow}  {e^{-}} W^{+}  b {\gamma}{\rightarrow} {e^{-}} \textit{l} {\nu}_\textit{l}  b {\gamma}   $   as a function of the anomalous ${\lambda}_{q}$ coupling for center-of mass energy of $\sqrt{s}=7.08$ TeV at the FCC-eh}\label{FIG2.}
\end{figure}

\begin{figure}[h]
  \centering
  \includegraphics[width=10.cm]{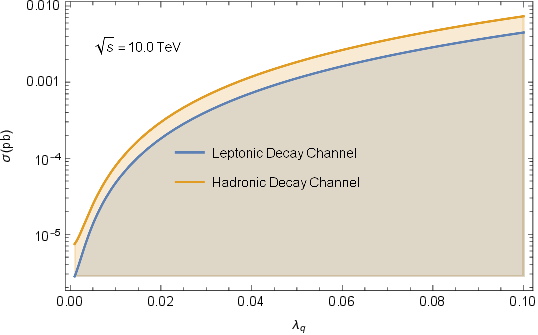}
   \caption{The total cross sections of the processes  ${e^{-}} {p} {\rightarrow} {e^{-}} {\gamma} {p}  {\rightarrow}  {e^{-}} W^{+}  b {\gamma}{\rightarrow} {e^{-}} {j} {\bar{j}}  b {\gamma}   $ and ${e^{-}} {p} {\rightarrow} {e^{-}} {\gamma} {p}  {\rightarrow}  {e^{-}} W^{+}  b {\gamma}{\rightarrow} {e^{-}} \textit{l} {\nu}_\textit{l}  b {\gamma}   $  as a function of the anomalous ${\lambda}_{q}$ coupling for center-of mass energy of $\sqrt{s}=10.0$ TeV at the FCC-eh}\label{FIG3.}
\end{figure}

In this part of the study, we performed statistical analysis through the ${e^{-}} {p} {\rightarrow} {e^{-}} {\gamma} {p}  {\rightarrow}  {e^{-}} W^{+}  b {\gamma}{\rightarrow} {e^{-}} {j} {\bar{j}}  b {\gamma}   $ and ${e^{-}} {p} {\rightarrow} {e^{-}} {\gamma} {p}  {\rightarrow}  {e^{-}} W^{+}  b {\gamma}{\rightarrow} {e^{-}} \textit{l} {\nu}_\textit{l}  b {\gamma}   $  processes with regard to examining sensitivity of anomalous ${\lambda}_q$ couplings. We have used two different statistical analysis method depending on the number of SM events. When the number of SM events is ${\leq}10$, the appropriate analysis method for this case is poisson analysis, while if the number of SM events $>10$ , the chi-square method is more suitable ones. In the Poisson analysis, the number of observed events is considered equal to the SM predictions, $N_{obs}=\mathcal{L}\times b_{tag} \times \sigma_{SM}=N_{SM}$. Here, $b_{tag}$ symbolizes tagging efficiency of b quark and it is chosen as $0.9$ in accordance with the experimental findings in the literature \cite{bquark1,bquark2}. And $\mathcal{L}$ is the integrated
luminosity. Here, we acquired upper limits of the number of events Nup at 95\% C.L. as follows \cite{63,64},

\begin{equation}
 \sum_{k=0}^{N_{obs}}P_{Poisson}(N_{up},k)=0.05 .
\end{equation}

On the other hand, at the 95\% C.L. the $\chi^{2}$ function in the $\chi^{2}$ criterion, with a systematic error is identified with following equation.
	\begin{eqnarray}
	{\chi^{2}=(\frac{\sigma_{SM}-\sigma_{Tot}}{\sigma_{SM}\sqrt{({\delta}_{stat})^2+({\delta}_{sys})^2}})^{2}},
	\end{eqnarray}
where, $\sigma_{SM}$ is the SM cross section,  $\sigma_{Tot}$ is the total cross section of process including both SM and new physics cross section. $ {\delta}_{stat}=\frac{1}{\sqrt{N_{SM}}} $ and $ {\delta}_{sys}$ show the statistical and the systematic error, respectively. Here, the $\chi^{2}$ function value at the 95\% C.L. is 3.84.

For sensitivity analysis, we have obtained constraints on the anomalous coupling parameter without systematic error. The results related to ${e^{-}} {p} {\rightarrow} {e^{-}} {\gamma} {p}  {\rightarrow}  {e^{-}} W^{+}  b {\gamma}{\rightarrow} {e^{-}} {j} {\bar{j}}  b {\gamma}   $ and ${e^{-}} {p} {\rightarrow} {e^{-}} {\gamma} {p}  {\rightarrow}  {e^{-}} W^{+}  b {\gamma}{\rightarrow} {e^{-}} \textit{l} {\nu}_\textit{l}  b {\gamma}   $  processes are presented in Table IV at centre-of -mass energy of 7.08 TeV for integrated luminosities 100, 500, 1000, 1500, 2000, 3000 $fb^{-1}$. Table V shows data on same processes and luminosities for $\sqrt{s}=10.0$ TeV. Moreover, we replicated our $\chi^{2}$ analysis for luminosity 2000 $fb^{-1}$ and 3000 $fb^{-1}$ in the hadronic channel at an assumption of 4\% systematic error.  $\lambda_{q}=0.002396$ for luminosity 2000 $fb^{-1}$ and $\lambda_{q}=0.002180$  for luminosity 3000 $fb^{-1}$ were found. These obtained results were compared with the obtained ones without systematic error (in the table V), it is seen that there is no a significant difference. For this reason, it can be inferred that the ${\delta}_{sys}=0$ selection does not cause a significant change in our results.

\begin{table}[h]
\caption{95\% C.L. bounds on the anomalous $\lambda_{q}$ coupling at the center-of-mass energy $\sqrt{s}=7.08$ TeV}
\begin{center}
\begin{tabular}{cc|cc|cc}
\midrule
\multicolumn{6}{c}{$\sqrt{s}=7.08 $ TeV} \\
\midrule
\multicolumn{2}{c}{} & \multicolumn{2}{c|}{Hadronic Channel} & \multicolumn{2}{c}{Leptonic Channel} \\
\midrule
${\cal L} \, (fb^{-1})$   &&
\hspace{1.5cm} $\lambda_{q}$ \hspace{1.5cm} &&
\hspace{1.5cm} $\lambda_{q}$ \hspace{1.5cm} & \\
\midrule
100   &&0.0091164    &&0.0138035&   \\
500   &&0.0046133    &&0.00580993&   \\
1000  &&0.00400351   &&0.00504988&   \\
1500  &&0.00331624   &&0.00389067&   \\
2000  &&0.00289991   &&0.00388222&  \\
3000  &&0.0026536    &&0.00336868&  \\
\botrule
\end{tabular}
\end{center}
\end{table}

\begin{table}
\caption{95\% C.L. bounds on the anomalous $\lambda_{q}$ coupling at the center-of-mass energy $\sqrt{s}=10.0$ TeV}
\begin{center}
\begin{tabular}{cc|cc|cc}
\toprule
\multicolumn{6}{c}{$\sqrt{s}=10.0 $ TeV} \\
\midrule
\multicolumn{2}{c}{} & \multicolumn{2}{c|}{Hadronic Channel} & \multicolumn{2}{c}{Leptonic Channel} \\
\midrule
${\cal L} \, (fb^{-1})$   &&
\hspace{1.5cm} $\lambda_q$ \hspace{1.5cm} &&
\hspace{1.5cm} $\lambda_q$ \hspace{1.5cm} & \\
\midrule
100   &&0.00798568  &&0.00832519&   \\
500   &&0.00384720  &&0.0042881&   \\
1000  &&0.00301846  &&0.00323541&   \\
1500  &&0.00264058  &&0.00276383&   \\
2000  &&0.00238525  &&0.00248182&  \\
3000  &&0.00216549  &&0.00214141&  \\
\botrule
\end{tabular}
\end{center}
\end{table}
Last part of the research is about the determination of the branching ratio limits on the anomalous $t\rightarrow q\gamma$ couplings. In these context, Figs 4 and 5 have been plotted $BR(t\rightarrow q\gamma)$  as a function of integrated luminosities via ${e^{-}} {p} {\rightarrow} {e^{-}} {\gamma} {p}  {\rightarrow}  {e^{-}} W^{+}  b {\gamma}{\rightarrow} {e^{-}} {j} {\bar{j}}  b {\gamma}   $ and ${e^{-}} {p} {\rightarrow} {e^{-}} {\gamma} {p}  {\rightarrow}  {e^{-}} W^{+}  b {\gamma}{\rightarrow} {e^{-}} \textit{l} {\nu}_\textit{l}  b {\gamma}   $  processes. In these two graphs, center-of-mass energy are accepted $\sqrt{s}=7.08 $ TeV and $\sqrt{s}=10.0$ TeV, respectively.

\begin{figure}
\includegraphics[width=12.cm]{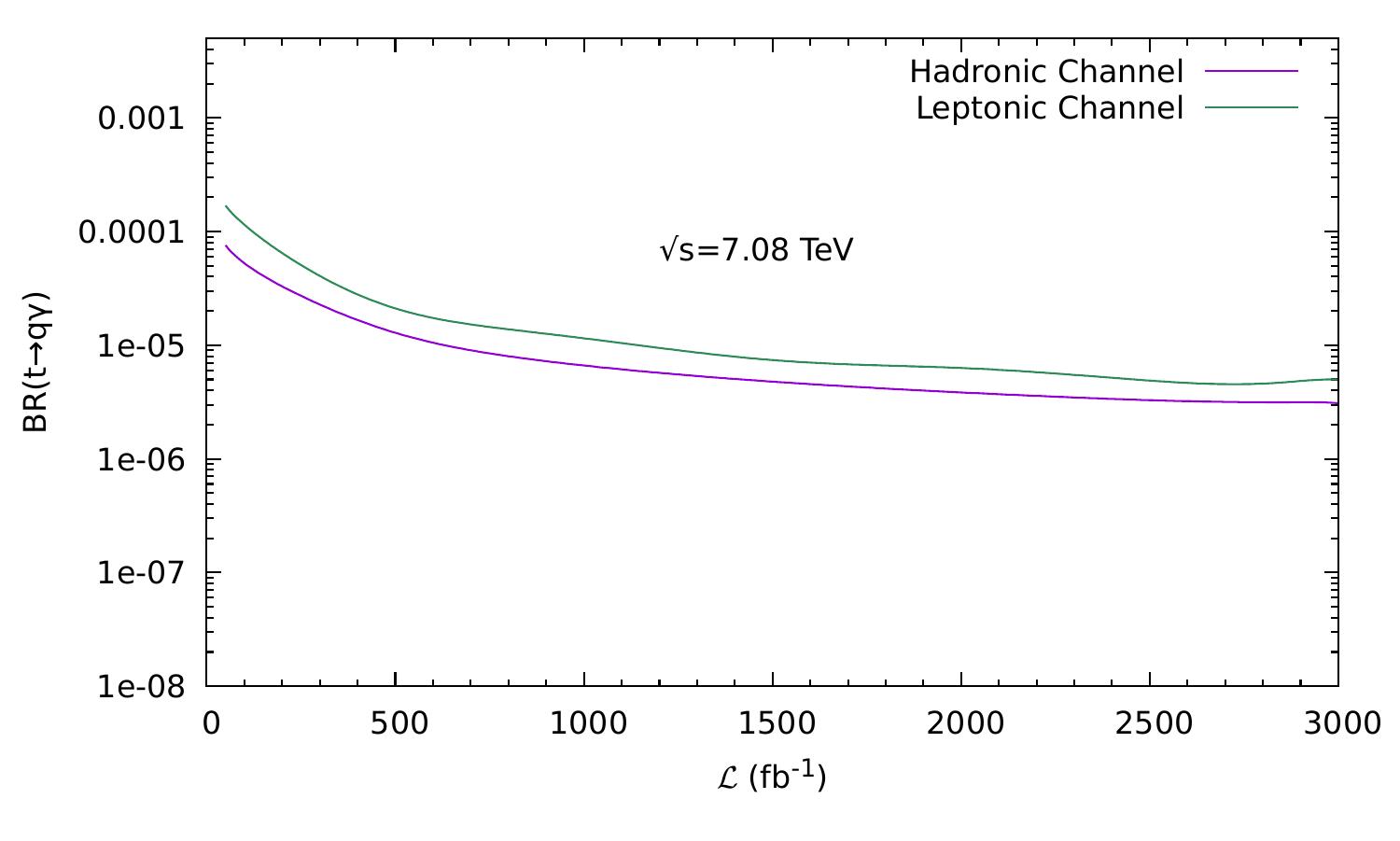}
\caption{For processes ${e^{-}}{p}{\rightarrow}{e^{-}}{\gamma}{p}{\rightarrow}{e^{-}}W^{+}b{\gamma}{\rightarrow}{e^{-}}{j} {\bar{j}}b{\gamma}$ and ${e^{-}} {p} {\rightarrow}{e^{-}}{\gamma}{p}{\rightarrow}{e^{-}}W^{+}b{\gamma}{\rightarrow}{e^{-}}\textit{l}{\nu}_\textit{l}b{\gamma} $, 95\% C.L. sensitivity limits on  $BR(t \rightarrow q\gamma)$
for various integrated luminosities at the FCC-eh}
\end{figure}

\begin{figure}
\includegraphics[width=12.cm]{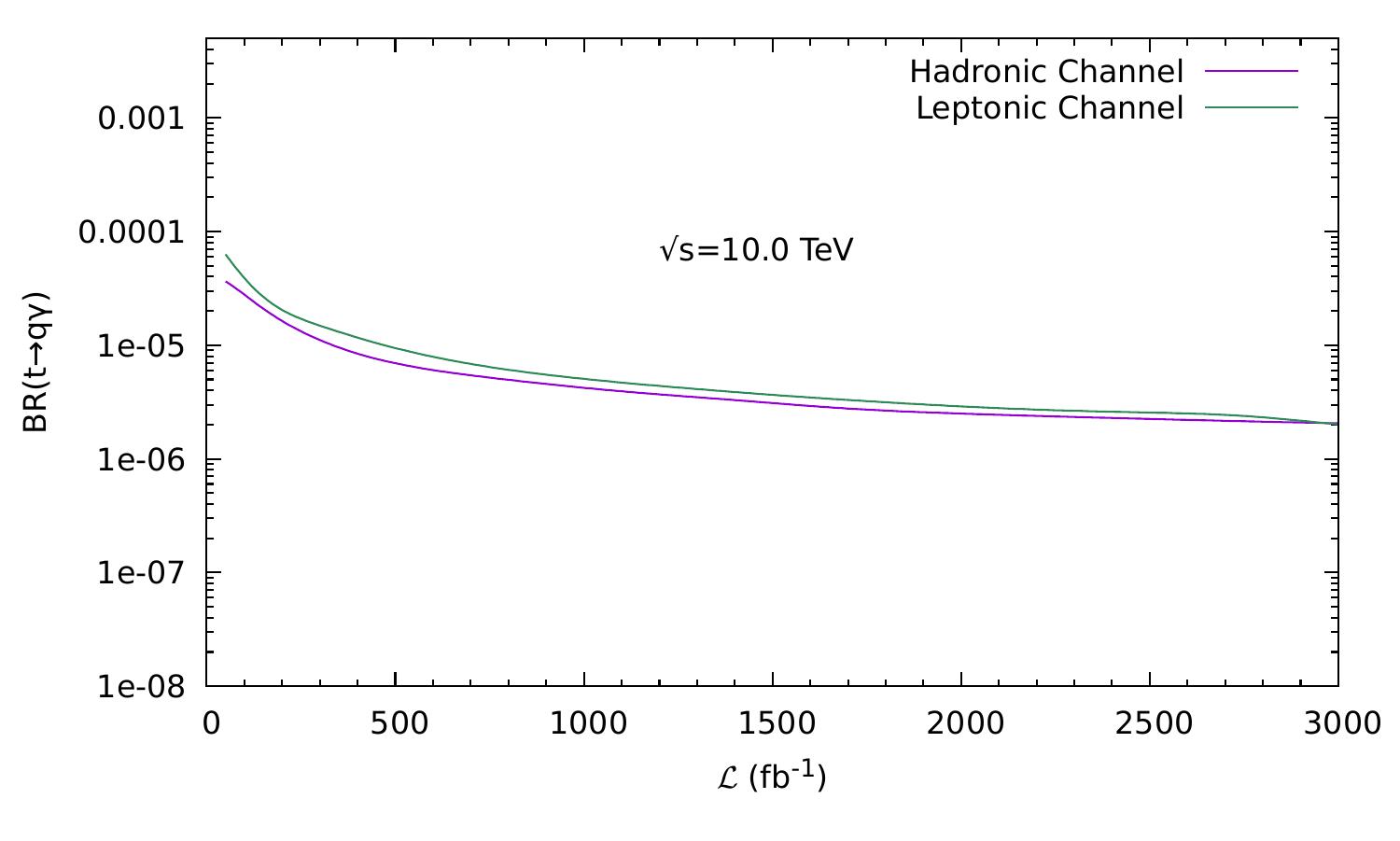}
\caption{Same as Fig.6, but for center-of mass energy of $\sqrt{s}=10.0$ TeV at the FCC-eh}
\end{figure}

According to the Figure 6 and Figure 7, it is realized that computed constraints on $BR(t\rightarrow     q\gamma)$ are better than the recent experimental data. In addition, for $\sqrt{s}=7.08$ TeV (as it can be seen from the Fig. 6), in $ \mathcal{L}=100 fb^{-1}$, the constraints was obtained by the ATLAS collaboration were improved 2 and 3 times in both leptonic and hadronic channels, respectively. Also, for luminosity 1500 $fb^{-1}$ value of the hadronic interaction process in the Figure 6, it was obtained $BR(t \rightarrow q\gamma)=4.81\times10^{-6}$, which corresponds 22 times more improved than the final experimental results. While The BR value obtained in luminosity  3000 $fb^{-1}$ for the hadronic channel are approximately 35 times better than the experimental result, this value is also 1.60 times stronger than BR value for the leptonic channel. Since the centre of mass energy in Fig. 7 ($\sqrt{s}=10.0$ TeV) is larger than the ones in Fig. 6 ($\sqrt{s}=7.08$ TeV),  better constraints were achieved on BR for $\sqrt{s}=10.0$ TeV. For the center-of mass energy of 10 TeV in the hadronic channel, the number of SM events  at $ \mathcal{L}{\geq}1600 fb^{-1}$ values is greater than 10. In this case, as mentioned before, two different analysis methods have been required. Hence, Poisson analysis was performed in the first part of the studies ($ \mathcal{L}{\leq}1600 fb^{-1}$) while for the second part, ($ \mathcal{L}{\geq}1600 fb^{-1}$), $\chi^{2}$ analysis was conducted. According to the Fig. 7, the hadronic channel was more restrictive on $BR(t\rightarrow     q\gamma)$ as same (Poisson) analysis up to luminosity $ 1600fb^{-1}$ was conducted for both hadronic and leptonic channel calculations. However, when the luminosity values greater than $ 1600fb^{-1}$, the difference between the BR limits get closer and almost equalizes because of the use of different analysis methods for the two channel. To exemplify this situation, while for $\mathcal{L}= 1000 fb^{-1}$, BR values are $4.59\times10^{-6}$ for the leptonic channel  and $3.99\times10^{-6}$ for the hadronic channel $\mathcal{L}= 3000 fb^{-1}$, BR values are $2.01\times10^{-6}$ for the leptonic channel and $2.05\times10^{-6}$ for the hadronic channel. In this context,both for the hadronic and leptonic channel at an centre-of mass energy 10 TeV, at $3000fb^{-1}$ in FCC-eh the strictest limits obtained in the study are conveyed in Fig. 6. This data is approximately 50 times stronger than the experimental limits reported by ATLAS Collaboration. In the light of all these findings, in this paper, the experimental limits on the anomalous photon, q branching of the top quark are restricted nearly by two orders of magnitude.

\section{Conclusion}\label{sec13}

One of the most focused areas by scientists in their researches on new physics beyond the SM is the examination of top quark interactions, the most massive particle of the SM. Also,  it is well established that an interesting subject of top quark physics is FCNC transitions. Although this type of transition is prohibited in SM, it is likely to be allowed in some of the BSM models. However, there is no evidence on FCNC transitions under current experimental conditions. Hence,it has been focused on future colliders in the recent phenomenological studies. In this context, our study are achieved the sensitivity limits for $tq\gamma$ coupling by considering ${e^{-}} {p} {\rightarrow} {e^{-}} {\gamma} {p}  {\rightarrow}  {e^{-}} W^{+}  b {\gamma}{\rightarrow} {e^{-}} {j} {\bar{j}}  b {\gamma}   $ and ${e^{-}} {p} {\rightarrow} {e^{-}} {\gamma} {p}  {\rightarrow}  {e^{-}} W^{+}  b {\gamma}{\rightarrow} {e^{-}} \textit{l} {\nu}_\textit{l}  b {\gamma}   $  from the $\gamma q $ processes that occur as a result of electron-proton interactions in the FCC collider, which is planned to be built in the future. Furhermore, our results were compared with the recent experimental limits reported by ATLAS collaboration and found more restrictive than ATLAS collaboration's ones. Here, the strictest limits were obtained for both the leptonic and hadronic processes at the $\sqrt{s}=10.0$ TeV and $ \mathcal{L}=3000 fb^{-1}$. These results indicate the improving of the experimental limits by 2 orders of magnitude. As a result, the future FCC-eh collider is promising for detecting the presence of anomalous FCNC transitions.

\section{Acknowledgments}
The author thanks to Prof. Dr. Murat Köksal from Sivas Cumhuriyet University who provided insights and expertise that greatly assisted the research.

\end{document}